\documentclass[aps,showpacs,prb,twocolumn]{revtex4}%
\usepackage{graphicx}
\usepackage{amsmath}
\usepackage{amsfonts}
\usepackage{amssymb}%
\begin{document}
\title{TRANSPORT PROPERTIES FOR AN INTERMEDIATE VALENCE MODEL OF Tl$_{2}$Mn$_{2}%
$O$_{7}$. }
\author{M.E. Foglio and G.E. Barberis}
\affiliation{Instituto de F\'{\i}sica \textquotedblright Gleb Wataghin\textquotedblright,
UNICAMP,13083-970, Campinas, S\~{a}o Paulo, Brazil.}

\begin{abstract}
The appearance of colossal magneto resistance (CM) in\textrm{ Tl}$_{2}%
$\textrm{Mn}$_{2}$\textrm{O}$_{7}$ has stimulated many recent studies of the
pyrochlore family of compounds \textrm{A}$_{2}$\textrm{B}$_{2}$\textrm{O}%
$_{7}$. The double exchange (DE) model of Zener does not describe
the CM in \textrm{ Tl}$_{2}$\textrm{Mn}$_{2}$\textrm{O}$_{7}$,
because its metallic conductivity cannot be explained by doping.
Here we employ Hubbard operators to reformulate the intermediate
valence model used by Ventura and Alascio to describe the
electronic structure and transport properties of this compound
$\left(  \mathit{Phys.Rev.}\ \mathbf{B}56,\ 14533\ (1997)\right)
$. Following Foglio and Figueira $\left(  \mathit{Phys.Rev.}\ \mathbf{B}%
62,\ 7882,\ (2000)\right)  $ we use approximate one-electron Green's Functions
(GF) to calculate the thermopower and the static and dynamic conductivity of
\textrm{Tl}$_{2}$\textrm{Mn}$_{2}$\textrm{O}$_{7}$ for several magnetic
fields. A qualitative agreement was obtained with the experimental
measurements of those properties. Although the agreement is far from perfect,
these quantities are fairly well described by the same set of system parameters.

\end{abstract}
\pacs{75.70.Pa.}
\date{\today }
\maketitle


\section{Introduction}

The A$_{2}$B$_{2}$O$_{7}$ pyrochlore family of compounds has been the object
of many recent studies, principally because of the appearance of colossal
magneto resistance (CM) in the \textrm{Tl}$_{2}$\textrm{Mn}$_{2}$%
\textrm{O}$_{7}$ compound. This material was reported as ferromagnetic,
metallic and with enormous negative magneto resistance in the region of
temperatures corresponding to the ferromagnetic transition (Tc $\sim$ 121 K).
The principal interest created for this material is that the double exchange
(DE) Zener model does not explain the CM in this material. Experimental
studies show that, when compared with other A$_{2}$Mn$_{2}$O$_{7}$ compounds,
the Tl one is unique. It has a very high Tc, a property shared with the In
pyrochlore, but it is the only compound of the series that presents metallic
conductivity, which cannot be explained by doping because it is present in the
stoichiometric compound. Several recent band calculations\cite{Ramirez1,
Mishra, Singh} show that the material is expected to be almost half metallic
at low temperatures, that is, the conductivity is driven within only one of
the spin directions. This property is the same for the DE perovskites, but in
that case the conductivity and the ferromagnetism appear only with doping.

The whole family of Mn pyrochlores shows ferromagnetism,\cite{Shimakawa2} and
their bands present similarities, with a band gap reducing its value from
Y\ to Tl, where the compound is slightly metallic with very different
conductivity for the both spin directions. The ferromagnetism in the Mn
pyrochlore compounds is explained by the Kanamori-Goodenough rules, as the
result of superexchange through the O ligands. However, the Curie temperatures
do not scale with the bond angles, which seems to indicate a different
mechanism for the In and Tl compounds\cite{Shimakawa2}(The calculated values
of the exchange parameter \textit{J} for the Tl, In, and Y compounds are
respectively 0.11 K, 2.52 K, and 1.1K, but their corresponding measured Tc's
are 124 K, 129 K and 16 K).

The experimental paper published by Raju et al.\cite{Raju}\ tries to explain
the ferromagnetism of the Tl compound through DE, but several evidences proved
afterwards that the origin of ferromagnetism is only superexchange.

A small number of carriers, of the order found in doped semiconductors
$[\sim10^{19}cm^{-3}]$, was found in the Tl compound by measuring Hall
effect\cite{Shimakawa1}$.$\emph{ }Several authors obtain a large enhancement
of the magnetoresistance\cite{Ramirez1,Martinez1} by doping the Tl sites with
In and Sc, or the Mn sites with Ru. The behavior of the resistivity is sample
dependent \cite{Shimakawa1, Cheong} at high T, which created a controversy
about the compound behavior in this region of temperatures.

Imai et al\cite{Imai} present the more complete set of measurements of Hall
effect and magneto-thermopower.\emph{ }They measured the anomalous Hall
coefficient at low (below Tc) and high temperatures, and they found it small
in both cases. Following Singh\cite{Singh} they assume a very simple \ quasi
spherical Fermi surface and fit the thermopower to the effective mass model.

The pressure reduces the value of Tc for all the Mn pyrochlores.\cite{Shusko}
Nu\~{n}ez- Regueiro and Lacroix\cite{Manolo} developed a theory for this
effect that gives good agreement with the experimental results, confirming
that the ferromagnetism is due to superexchange.

Ventura and Alascio\cite{Ventura1} used an intermediate valence (IV) model for
the Tl, and they could explain both the conductivity and the CM of the pure Tl
compound. Here we refine their calculations, using the same IV model, but
employing Hubbard X operators.\cite{Foglio1} We calculate both the static and
dynamic conductivity and a qualitative agreement with the measured quantities
was obtained\cite{Shimakawa0, Okamura}, without having to assume an O
deficiency. This agrees with the result of the published band
studies\cite{Ramirez1, Mishra, Singh}. Following Schweitzer and
Czycholl\cite{Schweitzer} we calculate the thermopower and magnetothermopower,
and our results agree with the experiments.\cite{Imai}

In Section II we reformulate the model of Ventura and Alascio, employing the
Hubbard operators. We discuss the approximate GF that we employ in the
calculation, and give the formulae to calculate the resistivity, optical
conductivity and thermopower. In Section III we \ discuss the parameters we
shall use, and calculate the transport properties and in Section IV we present
our conclussions. Finally, we give an Appendix with the atomic states employed
to calculate the GF.

\section{\label{}The model with $X$-operators}

The model employed \cite{Ventura1} is a lattice of local states hybridized
with a conduction band. \ Each local state has two magnetic configurations
with spin $S=1/2$ and $S=1$ respectively, that are hybridized with the
conduction electrons. The local state with $S_{z}=0$ of the $S=1$
configuration is discarded, so that we have two independent systems with spin
up and spin down respectively because the hybridization conserves the spin
direction. Each local site $j$ has then two states $\left\vert j,\sigma
\right\rangle $ with $S=\frac{1}{2},\sigma=\pm\frac{1}{2}$ and two states
$\left\vert j,s\right\rangle $ with $S=1,s=\pm1$, and their respective
energies are $E_{\sigma}$ and $E_{s}$.

We can give assume arbitrary properties and energies for the relevant
localized states of this model, and it is then convenient to describe them
employing Hubbard operators $X_{j;a,b}=\left\vert j,a\right\rangle
\left\langle j,b\right\vert $, which transform the state $\left\vert
j,b\right\rangle $ into the state $\left\vert j,a\right\rangle ,$ i.e.
$X_{j,ab}\left\vert j,b\right\rangle =\left\vert j,a\right\rangle $. These
operators do not satisfy Wick's theorem, and their product rules at the same
site have to be used instead:
\begin{equation}
X_{j,ab}\ X_{j,cd}=\delta_{b,c}\ X_{j,ad}. \label{E2.2}%
\end{equation}
When the operators are at different sites we chose properties equivalent to
those of the usual Fermi or Bose: we say that $X_{j,ab}$ is of the
\textquotedblleft Fermi type\textquotedblright\ (\textquotedblleft Bose
type\textquotedblright) when the number of electrons in the two states $\mid
j,a\rangle$ and $\mid j,b\rangle$ differ by an odd (even) number. For $j\neq
j^{\prime}$ we then use $\left\{  X_{j,ab},X_{j^{\prime},cd}\right\}  =0$ when
the two operators are of the \textquotedblleft Fermi type\textquotedblright%
\ and $\left[  X_{j,ab},X_{j^{\prime},cd}\right]  =0$ when at least one is of
the \textquotedblleft Bose type\textquotedblright\ (as usual\cite{FetterW}
$[a,b]=ab-ba$ and $\{a,b\}=ab+ba$).

We write the model's Hamiltonian employing :
\begin{align}
H  &  =\sum_{j\mathbf{,}\sigma}\ E_{\sigma}\ X_{j;\sigma\sigma}+\sum
_{j,s}E_{s}X_{j;ss}+\sum_{\mathbf{k},\sigma}E_{\mathbf{k},\sigma}%
c_{\mathbf{k},\sigma}^{\dagger}c_{\mathbf{k},\sigma}+\nonumber\\
&  +\sum_{j,\mathbf{k},\sigma}\left(  V_{j,\mathbf{k},\sigma}X_{j;\sigma
,s=2\sigma}^{\dagger}\ c_{\mathbf{k},\sigma}+V_{j,\mathbf{k},\sigma}^{\ast
}\ c_{\mathbf{k},\sigma}^{\dagger}X_{j;\sigma,s=2\sigma}\right)  \label{E2.3}%
\end{align}
where we denote the Hubbard operators $X_{j;\pm\frac{1}{2},\pm1}$ with
$X_{j;\sigma,s=2\sigma}$ (note that $X_{j;b,a}=X_{j;a,b}^{\dagger}$). The
$c_{\mathbf{k},\sigma}^{\dagger}$ and $c_{\mathbf{k},\sigma}$ are the creation
and destruction operators of a conduction electron with energies
$E_{\mathbf{k},\sigma}$, wave vector $\mathbf{k}$, and spin component
$\sigma\hbar/2$, where $\sigma=\pm1$, and the hybridization constant is%
\begin{equation}
V_{j,\mathbf{k},\sigma}=\left(  1/\sqrt{N_{s}}\right)  V(\mathbf{k}%
)\exp\left(  i\mathbf{k.{R}_{j}}\right)  , \label{E2.3a}%
\end{equation}
where $V(\mathbf{k})$ is independent of $\mathbf{k}$ when the mixing is purely
local and $N_{s}$ is the number of sites.

The cumulant expansion was extended by Hubbard\cite{Hubbard5} to study a
quantum problem with fermions, and he derived a diagrammatic expansion
involving unrestricted lattice sums of connected diagrams that satisfies a
linked cluster theorem. The extension of this technique to the Anderson
lattice\cite{FFM} is sufficiently general to treat the model described by Eq.
(\ref{E2.3}), and is the basis of the present treatment. One has to use the
Grand Canonical Ensemble of electrons, and it is then convenient to introduce
\begin{equation}
\mathcal{H}=H-\mu\left\{  \sum_{\vec{k},\sigma}C_{\vec{k},\sigma}^{\dagger
}C_{\vec{k},\sigma}+\sum_{ja}\nu_{a}X_{j,aa}\right\}  \qquad, \label{E2.6}%
\end{equation}

\noindent where $\mu$ is the chemical potential and $\nu_{a}$ is the number of
electrons in the state $\left\vert j,a\right\rangle $, and without any
restriction for the treatment we shall use $\nu_{\pm\frac{1}{2}}=0$ and
$\nu_{\pm1}=1$. It is also convenient to introduce
\begin{align}
\varepsilon_{j,a}  &  =E_{j,a}-\mu\ \nu_{a}\label{E2.6a}\\
\varepsilon_{\mathbf{k\sigma}}  &  =E_{\mathbf{k\sigma}}-\mu\ \qquad,
\label{E2.6b}%
\end{align}
because these are the forms that consistently appear in the calculations.

The last term in Eq. (\ref{E2.3}) will be considered as the perturbation, and
the exact and unperturbed averages of any operator $A$ shall be respectively
denoted by $<A>_{\mathcal{H}}$ and $<A>$.

\subsection{The Approximate Green's functions\textbf{\ }}

As in the Anderson lattice\cite{Foglio1} with $U\rightarrow\infty$ one can
introduce one-particle Green's functions (GFs) of local electrons
\begin{equation}
\left\langle \left(  X_{j;\sigma,s=2\sigma}(\tau)\ X_{j;\sigma,s=2\sigma
}^{\dagger}(\tau^{\prime})\right)  _{+}\right\rangle _{\mathcal{H}}\qquad,
\label{E4.8}%
\end{equation}
as well as GFs\ for the c-electrons $\left\langle \left(  C_{k\sigma}%
(\tau)\ C_{k^{\prime}\sigma}^{\dagger}(\tau^{\prime})\right)  _{+}%
\right\rangle _{\mathcal{H}}$\ and \textquotedblleft crossed\textquotedblright%
\ GFs of the type $\left\langle \left(  X_{j;\sigma,s=2\sigma}(\tau
)\ C_{k^{\prime}\sigma}^{\dagger}(\tau^{\prime})\right)  _{+}\right\rangle
_{\mathcal{H}}$, all of them defined in the intervals $0\leq\tau,\tau^{\prime
}\leq\beta\equiv1/T$. It is possible to associate a Fourier series to these
GFs because of their boundary condition in this variable,\cite{FFM} and the
coefficients $\left\langle \left(  X_{j;\sigma,s=2\sigma}(\omega_{\nu
})\ X_{j;\sigma,s=2\sigma}^{\dagger}(\omega_{\nu^{\prime}})\right)
_{+}\right\rangle _{\mathcal{H}+}$ correspond to the Matsubara frequencies
$\omega_{\nu}=\pi\nu/\beta$ (where $\nu$ are all the positive and negative odd
integer numbers). One can also transform the GF to reciprocal space,\cite{FFM}
and because of the invariance against time and lattice translations
\begin{align}
&  \left\langle \left(  X_{j;\sigma,s=2\sigma}(\omega_{\nu})\ X_{j;\sigma
,s=2\sigma}^{\dagger}(\omega_{\nu^{\prime}})\right)  _{+}\right\rangle
_{\mathcal{H}}=\nonumber\\
&  G_{ff,\sigma}(\mathbf{k},\omega_{\nu})\ \delta_{\mathbf{k}^{\prime
},\mathbf{k}}\ \delta_{\nu+{\nu}^{\prime},0}. \label{E4.8b}%
\end{align}
Transforming the eigenstates of the c-electrons to the Wannier representation,
one also obtains the equivalent relations for $G_{cc,\sigma}(\mathbf{k}%
,\omega_{\nu})$ and $G_{fc,\sigma}(\mathbf{k},\omega_{\nu})$. Considering that
the coefficients of the $\tau$ Fourier series for each $\mathbf{k}$ are the
values of a function of the complex variable $z=\omega+iy$ at the points
$z_{\nu}=i\ \omega_{\nu}$, it is possible to make the analytic continuation to
the upper and lower half-planes of $z$ in the usual way,\cite{Negele}
obtaining, e.g. from the $G_{ff,\sigma}(\mathbf{k},\omega_{\nu})$, a function
$G_{ff,\sigma}(\mathbf{k},z)$ which is minus the Fourier transform of the
double time GF.\cite{Zubarev}

The one-electron GF of ordinary Fermions or Bosons can be expressed as a sum
of infinite \textquotedblleft proper\textquotedblright\ (or irreducible)
diagrams,\cite{FetterW} and a similar result was obtained for the Hubbard
model employing the cumulant expansion\cite{Metzner} with the hopping as
perturbation. In the cumulant expansion of the Anderson lattice\cite{FFM} we
employed the hybridization rather than the hopping as a perturbation, and the
exact solution of the conduction electrons problem in the absence of
hybridization was included in the zeroth order Hamiltonian. It was then
necessary to extend Metzner's derivation\cite{Metzner} to the Anderson
lattice, and the same type of results he derived were also obtained for the
Anderson lattice. As with the Feynmann diagrams, one can rearrange all those
that contribute to the exact $G_{ff,\sigma}(\mathbf{k},\omega_{\nu})$ by
defining an effective cumulant $M_{2,\sigma}^{eff}(\mathbf{k},\omega_{\nu})$,
that is given by all the diagrams of $G_{ff,\sigma}(\mathbf{k},\omega_{\nu})$
that can not be separated by cutting a single edge (usually called
\textquotedblleft proper\textquotedblright\ or \textquotedblleft
irreducible\textquotedblright\ diagrams). The exact one-particle GFs of the
Anderson lattice\cite{Foglio1,Foglio2} \ were then obtained by introducing the
$M_{2,\sigma}^{eff}(\mathbf{k},\omega_{\nu})$ in the cumulant expansion, and
the model employed in those works was sufficiently general, so that their
results could be easily extended to the Hamiltonian in Eq. (\ref{E2.3}) in the
present work.

By analytical continuation one then obtains the formal expressions of the
exact one-particle GFs of our model:%
\begin{equation}
G_{ff,\sigma}(\mathbf{k},z)=\frac{M_{2,\sigma}^{eff}(\mathbf{k}{,}z)}{1-\mid
V(\mathbf{k})\mid^{2}G_{c{,}\sigma}^{o}(\mathbf{k}{,}z)\ M_{2,\sigma}%
^{eff}(\mathbf{k}{,}z)}, \label{E2.8}%
\end{equation}
and
\begin{equation}
G_{cc,\sigma}(\mathbf{k},z)=\frac{-\ 1}{z-\varepsilon_{\mathbf{k\sigma}}+\mid
V(\mathbf{k})\mid^{2}M_{2,\sigma}^{eff}(\mathbf{k}{,}z)}, \label{E2.8b}%
\end{equation}
where $G_{c{,}\sigma}^{o}(\mathbf{k}{,}z)=-1/(z-\varepsilon_{\mathbf{k\sigma}%
})$ is the free c-electron propagator.

The calculation of $\ M_{2,\sigma}^{eff}(\mathbf{k}{,}\omega_{\nu})$ is as
difficult as that of $G_{ff,\sigma}(\mathbf{k},\omega_{\nu})$, and it is then
convenient to use an approximation: we shall replace$\ M_{2,\sigma}%
^{eff}(\mathbf{k}{,}\omega_{\nu})$ by the corresponding quantity $M_{2,\sigma
}^{at}(\omega_{\nu})$ of an exactly soluble Hamiltonian, namely the one
describing the atomic limit of the same model. Although the hopping is
neglected in this system, described by the Hamiltonian of Eqs. (\ref{E2.3})
with $E_{\mathbf{k},\sigma}=E_{0,\sigma}$ and with a local hybridization
$V(\mathbf{k})=V$, the $M_{2,\sigma}^{at}(\omega_{\nu})$ implicitly contains
all the higher order cumulants that appear in the exact quantity. In the case
of the Anderson lattice, the atomic limit contains the basic physics of the
formation of the singlet ground state and of the appearance of the Kondo
peak,\cite{Atomic1,Atomic2} and we expect that it would provide an adequate
description of the present model. Because of its atomic character, the
approximate effective cumulant $M_{2,\sigma}^{at}(\omega_{\nu})$ thus obtained
is independent of $\mathbf{k}$,\ and can be calculated exactly as discussed below.

With the approximations introduced above and employing the Wannier
representation for the $c$-electron operators the whole system becomes a
collection of local systems, described by a Hamiltonian $\sum_{j}H_{j}$, where
$H_{j}$ \ is the local Hamiltonian at site $j$. This $H_{j}$ can be solved
exactly:%
\begin{equation}
H_{j}\mid j,\nu,r\rangle=E_{\nu,r}\mid j,\nu,r\rangle, \label{E2.9}%
\end{equation}
where $\mid j,\nu,r\rangle$ is the eigenstate at site $j$ with energy
$E_{\nu,r}$, that is characterized by $r$ and its number $\nu$ of electrons.
Because of the translational invariance we shall drop the site index $j$ when
it is not necessary, and we shall also use the quantities $\varepsilon_{\nu
,r}=E_{\nu,r}-\nu\mu$, more adequate for the $\mathcal{H}$ in Eq. (\ref{E2.6})
than the $E_{\nu,r}$ (for convenience we use $\nu_{\pm\frac{1}{2}}=0$ and
$\nu_{\pm1}=1$). In the Appendix we give in Table \ref{T1} the properties of
the $\mid\nu,r\rangle$ states: the number $r$ that identifies the state, the
$z$ component of spin $S_{z}$ and the quantities $\varepsilon_{\nu,r}%
=E_{\nu,r}-\nu\mu$ .\

It is now straightforward to express the Fourier transform $G_{ff,\sigma}%
^{at}(\omega_{s})$ of the f-electron GF in the atomic limit
\begin{align}
G_{ff,\sigma}^{at}(\omega_{s})  &  =-e^{\beta\Omega}\sum_{\nu,r,r^{\prime}%
}\frac{\exp(-\beta\varepsilon_{\nu,r})+\exp(-\beta\varepsilon_{\nu
-1,r^{\prime}})}{i\omega_{s}+\varepsilon_{\nu-1,r^{\prime}}-\varepsilon
_{\nu,r}}\nonumber\\
\times &  \mid\langle\nu-1,r^{\prime}\mid X_{\sigma,s=2\sigma}\mid\nu
,r\rangle\mid^{2}, \label{E2.12}%
\end{align}
where $\Omega=-kT\ln\sum\exp(-\beta\epsilon_{\nu,r})$ is the grand canonical
potential.\cite{Martinez} The equivalent equations for the c-electrons are
obtained by just replacing $\mid\langle\nu-1,r^{\prime}\mid X_{\sigma
,s=2\sigma}\mid\nu,r\rangle\mid^{2}$ in Eq.~(\ref{E2.12}) by $\mid\langle
\nu-1,r^{\prime}\mid C_{j,\sigma}\mid\nu,r\rangle\mid^{2}$.

The f-electron GF can be written in the form%

\begin{equation}
G_{ff,\sigma}^{at}(\omega_{s})=-\exp(\beta\Omega)\ \sum_{j=1}^{8}\frac{m_{j}%
}{i\omega_{s}-u_{j}}\qquad, \label{E2.13}%
\end{equation}
and the poles $u_{i}$ and residues $m_{i}$ of $G_{ff,\sigma}^{at}(\omega_{s})$
are all real (cf. Eq. (\ref{E2.12})). There are only eight different $u_{j}$
for the f-electron GF, because different transitions have the same energy and
the residues of some transitions are zero, and by analytic continuation one
obtains $G_{ff,\sigma}^{at}(z)$, but there are more transitions for the
$G_{cc,\sigma}^{at}(z)$.

The approximation employed in the present work consists in substituting
$M_{2,\sigma}^{eff}(z)$ in Eq.~(\ref{E2.8}) by the approximate $M_{2,\sigma
}^{at}(z)$, derived from the exact $G_{ff,\sigma}^{at}(z)$ by solving for
$M_{2,\sigma}^{at}(z)$ in the equation that is the atomic equivalent of
Eq.~(\ref{E2.8}). One then obtains
\begin{equation}
M_{2,\sigma}^{at}(z)=\ \frac{\left(  z-E_{0}^{a}+\mu\right)  \ G_{ff,\sigma
}^{at}(z)}{\left(  z-E_{0,\sigma}^{a}+\mu\right)  -\mid V\mid^{2}G_{ff,\sigma
}^{at}(z)}, \label{E2.14}%
\end{equation}
and from the point of view of the cumulant expansion, it contains all the
irreducible diagrams that contribute to the exact $M_{2,\sigma}^{eff}%
(\omega_{s})$. It should be emphasized that this diagrams contain loops of any
size, because there is no excluded site in this expansion, but all the local
vertices correspond to the same site, although they appear as different
vertices in each diagram. When a local hybridization is used (i.e.
$V(\mathbf{k})=V$), the only difference between the exact and approximate
quantities is that different energies $E_{\mathbf{k},\sigma}$ appear in the
c-electron propagators of the effective cumulant $M_{2,\sigma}^{eff}%
(\omega_{s})$, while these energies are all equal to $E_{0,\sigma}^{a}$ in
$M_{2,\sigma}^{at}(\omega_{s})$. Although $M_{2,\sigma}^{at}(\omega_{s})$ is
for that reason only an approximation, it contains all the diagrams that
should be present, and one would expect that the corresponding GF would have
fairly realistic features.

One still has to decide what value of $E_{0,\sigma}^{a}$ should be taken. As
the most important region of the conduction electrons is the Fermi energy, we
shall use $E_{0,\sigma}^{a}=\mu-\delta E_{0}$, leaving the freedom of small
changes $\delta E_{0}$ to adjust the results to particular situations, but
fixing its value for a given system when $\mu$ has to change to keep the total
number of electrons $N_{t}$ fixed, as for example when changing the
temperature $T$.

Another important point, is that concentrating all the conduction electrons at
$E_{0,\sigma}^{a}$ would overestimate their contribution to the effective
cumulant, and we shall then reduce the hybridization by a coefficient that
gives the fraction of c-electrons which contribute most. We consider that this
is of the order of $V\varrho^{0}$, where $\varrho^{0}$ is the density of
states of the free c electrons per site and per spin, and to be more definite
we chose $\pi V\varrho^{0}$, so the effective hybridization constant $V_{a}$
coincides with the usual \textquotedblleft mixing strength\textquotedblright%
\ $\Delta=\pi V^{2}\varrho^{0}$. This is essentially the same choice made by
Alascio et.al.\cite{AlascioAA} in their localized description of valence
fluctuations. Note that $V_{a}$ is only used in the calculation of
$M_{2,\sigma}^{at}(z)$, and that the full value must be substituted in the $V$
that appears explicitly in Eq. (\ref{E2.8}), because the whole band of
conduction energies is used in $G_{c{,}\sigma}^{o}({\vec{k},}z)=-1/\left(
z-\varepsilon_{\mathbf{k\sigma}}\right)  $.

\subsection{Transport properties}

\label{S03B}

Two particle GF should be used in the well known Kubo
formula,\cite{Kubo1,Mahan} that relates the dynamic conductivity
$\sigma\left(  \omega,T\right)  $ to the current current correlations. To
simplify the calculations for the PAM, Schweitzer and
Czycholl\cite{SchweitzerC} employed the expression of the conductivity for
dimension $d=\infty$ as an approximation of the static conductivity for $d=3$.
Only one-particle GFs are then necessary to obtain $\sigma\left(
\omega,T\right)  $ in that limit, because the vertex corrections cancel
out,\cite{Khurana} and we shall use the same approximation. As the
hybridization is a hopping of electrons between two different bands, it
contributes to the current operator,\cite{CzychollL} but this contribution
cancels out in our model because we employ a local hybridization
$V_{j,\mathbf{k},\sigma}=V_{j,\sigma}$. The expression obtained contains
explicit sums over $\mathbf{k}$, but it is possible to make a further
simplification by considering nearest-neighbor hopping in a simple cubic
lattice,\cite{MutouH,PruschkeCJ,ConsiglioG} and the sums over $\mathbf{k}%
$\ can be transformed\cite{MullerH} in integrals over the free conduction
electron energy $\varepsilon\left(  \mathbf{k}\right)  $. This transformation
is possible because in our method the $G_{cc,\sigma}(\mathbf{k},\omega)$ only
depends on $\mathbf{k}$ through the $\varepsilon(\mathbf{k})=\varepsilon$, as
both $M_{2,\sigma}^{at}(z)$ and $V_{j,\mathbf{k},\sigma}=V_{j,\sigma}$ are
$\mathbf{k}$ independent. We then obtain for the dynamic conductivity for each
spin component%
\begin{equation}
\sigma_{\sigma}\left(  \omega,T\right)  =C_{0}\frac{1}{\omega}\int_{-\infty
}^{\infty}d\omega\prime\left[  f_{T}(\omega\prime)-f_{T}(\omega\prime
+\omega)\right]  \ L_{\sigma}\left(  \omega,\omega^{\prime}\right)
\label{E3.1}%
\end{equation}
where%
\begin{equation}
L_{\sigma}\left(  \omega,\omega^{\prime}\right)  =\int_{-\infty}^{\infty
}d\varepsilon\ \rho_{c,\sigma}(\omega\prime;\varepsilon)\ \rho_{c,\sigma
}(\omega\prime+\omega;\varepsilon)\ \varrho_{\sigma}^{0}(\varepsilon),
\label{E3.2}%
\end{equation}%
\begin{equation}
\rho_{c,\sigma}(\omega;\varepsilon)=\frac{1}{\pi}\ \lim_{\eta\rightarrow
0}\ \mathit{Im}\left\{  G_{cc,\sigma}(\mathbf{k},\omega+i\left\vert
\eta\right\vert )\right\}  , \label{E3.3}%
\end{equation}
and $f_{T}(\omega)$ is the Fermi function. The static conductivity for each
spin component is then given by
\begin{equation}
\sigma_{\sigma}\left(  T\right)  =C_{0}\int_{-\infty}^{\infty}d\omega\ \left(
-\ \frac{df_{T}(\omega)}{d\omega}\right)  \ L_{\sigma}(\omega)\ \qquad,
\label{E3.4}%
\end{equation}
where
\begin{equation}
L_{\sigma}(\omega)=\int_{-\infty}^{\infty}d\varepsilon\ \left(  \rho
_{c,\sigma}(\omega;\varepsilon)\right)  ^{2}\ \varrho_{\sigma}^{0}%
(\varepsilon)\ \qquad. \label{E3.4a}%
\end{equation}

The constant
\begin{equation}
C_{0}=\pi\ \frac{e^{2}}{\hbar}\ \frac{2}{a}\ \frac{2\ t^{2}}{d}, \label{E3.6a}%
\end{equation}
where $a=9.89\mathring{A}$\ is the lattice parameter\cite{Shimakawa2} of
\textrm{Tl}$_{2}$\textrm{Mn}$_{2}$\textrm{O}$_{7}$, that has two sites per
unit cell. We shall generally use a rectangular band with $-W\leqslant
\varepsilon\left(  \mathbf{k}\right)  $ $\leqslant W$, and we set $t=W/2d$ to
estimate the hopping parameter $t$ of the hypercubic lattice, and use $d=3$.

Employing reference [\onlinecite{SchweitzerC}] we obtain the expression for
the thermopower $S(T)$ :%
\begin{equation}
S(T)=\frac{\sum_{\sigma}\int_{-\infty}^{\infty}d\omega\ \omega\ \left(
-\ df_{T}(\omega)/d\omega\right)  L_{\sigma}\left(  \omega\right)  }%
{\sum_{\sigma}\int_{-\infty}^{\infty}d\omega\ \left(  -\ df_{T}(\omega
)/d\omega\right)  L_{\sigma}\left(  \omega\right)  } \label{E3.5}%
\end{equation}

\subsection{The magnetization of the system}

As the system consists of two independent subsystems (spin up and spin down)
we could attribute arbitrary probabilities $P$ ($1-P$) to the spin up (spin
down) subsystem, and calculate the corresponding properties of the system.
Following the work of Ventura and Alascio, we shall estimate the probability
$P$ from the system magnetization, that we calculated employing the Weiss
molecular field approximation:
\[
\frac{M}{M_{sat}}=\tanh\left\{  \frac{\widetilde{\mu}\ B}{k_{B}\ T}%
+\frac{T_{C}\ M}{T\ M_{sat}}\right\}  .
\]
Here $\widetilde{\mu}$ is the local magnetic moment, $B$ the magnetic field,
$T_{C}$ the Curie temperature and $M_{sat}$ the saturation magnetization.
Following those authors we use $\widetilde{\mu}=3\mu_{B}$, as intermediate
between the $3.87\ \mu_{B}\ $local moment of the \textrm{Mn}$^{5+}$ and the
$2.83\ \mu_{B}$ corresponding to the \textrm{Mn}$^{4+}$.

To calculate the probability $P$ we then employ
\[
M=\left[  P-(1-P)\right]  \ \widetilde{\mu},
\]
and proceed to calculate the system properties as a function of $T$ for
different values of the total number $n$ of electrons per site. Employing our
approximate GF it is possible to calculate $n$ for each value of $P$ and $T$,
and it is necessary to find the chemical potential $\mu$ that gives the
required number of electrons per site.

\section{The calculation of the transport properties}

We shall consider the stoichiometric \textrm{Tl}$_{2}$\textrm{Mn}$_{2}%
$\textrm{O}$_{7}$ compound, and we shall then fix the total number of
electrons per site as $n=1$. To keep this value constant, it might be
necessary to change the chemical potential $\mu$ with the temperature $T$, and
we shall employ the approximate GF $G_{ff,\sigma}^{at}(z)$ and $G_{cc,\sigma
}^{at}(z)$ to calculate the number $n$ at each $T$ and then solve numerically
the equation $n=1$.

We shall use a rectangular band centered at the energy origin and with a half
width $W=6$ eV, and take the energies for the spin $1/2$ and $1$ in the
presence of the field $B$ as $E_{s}$ = $2s\ \mu_{B}\ B$ and $E_{\sigma}$ =
$E_{\sigma}^{0}+\sigma\ 2\mu_{B}\ B$ , with $E_{\sigma}^{0}=-5.5$ eV, so that
the spin $1/2$ has the lowest energy of the local states at $B=0$.

It seems clear that the basic scattering mechanism in our calculation of
$\sigma\left(  T\right)  $ is the hybridization, because the otherwise free
conduction electrons are scattered by the localized f electrons through this
interaction. This is apparent if we notice that the relaxation effects are
described by the imaginary part of the usual self-energy $\Sigma_{cc,\sigma
}(\mathbf{k,}z)$, defined through
\begin{equation}
G_{cc,\sigma}(\mathbf{k},z)=-\left\{  z-\varepsilon(\mathbf{k})-\Sigma
_{cc,\sigma}(\mathbf{k,}z)\right\}  ^{-1}\qquad, \label{E4.1}%
\end{equation}
and that the exact relation $\Sigma_{cc,\sigma}(\mathbf{k,}z)=-\mid
V(\mathbf{k})\mid^{2}M_{2,\sigma}^{eff}(\mathbf{k}{,}z)$ follows from Eq.
(\ref{E2.8b}). The relaxation mechanism of the c-electrons is then provided by
the hybridization, and in our approximation the self energy $\Sigma
_{cc,\sigma}(z)=-\mid V\mid^{2}M_{2,\sigma}^{at}(z)$ is independent of
$\mathbf{k}$. It seems then clear that the resistivity at low temperatures
depends sensitively on the value of $V$. This quantity also determines the
position of the peak of the dynamic conductivity that is close to $2$ e.V at
$295$ K, and to try and adjust the two different properties we have employed a
temperature dependent hybridization $V$, using values that decrease from $2.5$
eV to $1.8$ eV as $T$ increases to $300$ K.

We have also employed a value of $\Delta E_{0}=E_{0}^{a}-\mu$ that changes
from $-1.2$ eV to $-0.9$ eV in the same temperature range, because it gives a
better overall agreement with the experimental results.

To alleviate somehow the use of a zeroth width conduction band in the
calculation of the effective cumulant we have added an extra imaginary part
$\eta_{a}$ $=0.18$ eV to the complex variable $z$. Addition of $\eta_{a}$ to
the argument of $M_{2,\sigma}^{at}(z)$ leads to similar effects as those
already obtained by Mutou and Hirashima\cite{MutouH} through \textquotedblleft
introducing a small imaginary part $\Gamma$ to the conduction
electrons\textquotedblright, i.e. replacing $z=i\ \omega$ by $z+i\ \Gamma
\ sgn(\omega)$ in the GFs $G_{ff,\sigma}(\mathbf{k},z)$ and $G_{cc,\sigma
}(\mathbf{k},z)$. Their justification is the existence in real systems of
scattering processes due to phonons and impurities, and we should also
consider this mechanisms as contributing to the $i\eta_{a}$. Within this
interpretation one could also consider a temperature dependence of $\eta_{a}$,
but we have not implemented this change in the present calculation.

\subsection{The local spectral density of states}

\begin{figure}[tbh]
\includegraphics*[scale=0.80]{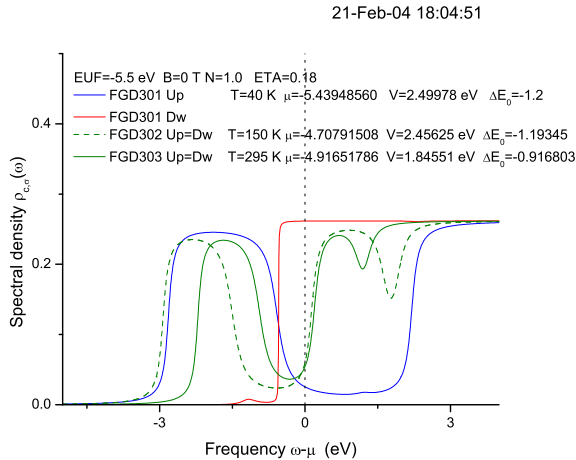}\caption{The local spectral
density of the conduction electrons $\rho_{c,\sigma}(\omega)$ at $T$ both
below and above the critical temperature $T_{C}=121$K, and for system
parameters indicated in the figure. We employ a $T$ dependent $V$ and $\Delta
E_{0}$, and their values together with $\mu$ are shown for each curve.}%
\label{Fg01}%
\end{figure}

\bigskip A very useful quantity is the local spectral density of the
conduction electrons, namely
\begin{equation}
\rho_{c,\sigma}(\omega)=\frac{1}{\pi}\ \lim_{\eta\rightarrow0}\ \mathit{Im}%
\left\{  \frac{1}{N_{s}}\sum_{\mathbf{k}}G_{cc,\sigma}(\mathbf{k}%
,\omega+i\ \left\vert \eta\right\vert )\right\}  \label{E4.2}%
\end{equation}
because it illuminates the dependence with $T$ of the static conductivity. In
figure \ref{Fg01} we plot $\rho_{c,\sigma}(\omega)$ at several temperatures
above and below the critical temperature \ $T_{C}=121$K, and for parameters
that give a fair description of the properties we study.

At $T=40$ K the $\rho_{c,\sigma}(\omega)$ is different for the two spin
components. The magnetization is practically saturated and all the local spins
point in the same direction, say up. The conduction electrons with spin up
hybridize with the local spins, and in figure \ref{Fg01} is shown that a large
gap is created with $\mu$ inside, so there is little conductivity by these
electrons. As there are practically no local electrons with spin down, the
conduction electrons do not have electrons to hybridize with, and there is no
gap. The chemical potential $\mu$ is near near the bottom of the band, and the
spin down electrons contribute strongly to the conductivity giving a vanishing
resistivity as it is shown in figure \ref{Fg02}.

At $T=295$ K the probabilities of the two spin components at zero magnetic
field are equal, and have the same spectral densities as shown in figure
\ref{Fg01}. The $\mu$ is inside a smaller gap, and the resistivity has
increased, as shown in figure \ref{Fg02}. \ For $T=150$ K the $\rho_{c,\sigma
}(\omega)$ has a larger gap but essentially the same features shown by the
plot at $T=295$ K, although the hybridization and the $\Delta E_{0}$ employed
here are practically the same used in the plot at $T=40$ K.

\begin{figure}[tbh]
\includegraphics*[scale=0.80]{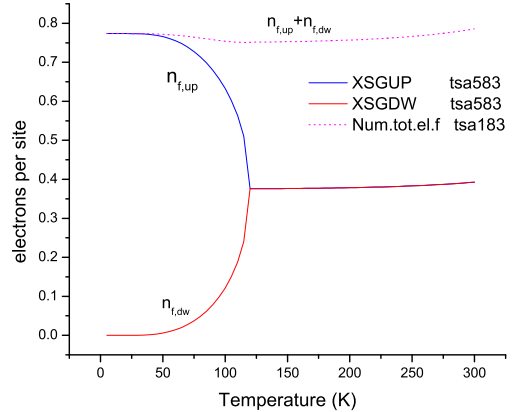}\caption{The number of local
electrons per site for spin up, spin down and for their sum, as a function of
$T$. The system parameters are indicated in the figure. }%
\label{Fg09}%
\end{figure}

Integrating the spectral densities over $\omega$ we obtain the number of up
($n_{f,up}$) and down ($n_{f,dw})$ local electrons at each $T$, and we plot
their values in figure \ref{Fg09}. This sum of these two quantities gives the
total number ($n_{f}$) of local electrons, and the plot shows that this
quantity is fairly independent of $T$. The measurements of X ray absorption
spectroscopy (XAS) in \textrm{Tl}$_{2}$\textrm{Mn}$_{2}$\textrm{O}$_{7}%
$\cite{Booth} indicate that the \textrm{Mn }valence is fairly close to 4. Two
facts point to the independence of this quantity with temperature. First, the
XAFS indicates that the local structure coincides with the average one in
\textrm{Tl}$_{2}$\textrm{Mn}$_{2}$\textrm{O}$_{7}$ and that there is no
disorder in the same structures of this compound, differently from the
disorder in the \textrm{MnO}$_{6}$ octahedra of the manganites, that is caused
by Jahn-Teller distortions. Second, the \textrm{Mn-O} and\textrm{\ Tl-O} bonds
in \textrm{Tl}$_{2}$\textrm{Mn}$_{2}$\textrm{O}$_{7}$ show normal Debye-like
dependence, with no change in ordering at $T_{c}$, in contrast with the
behavior of the \textrm{La}$_{0.75}$\textrm{Ca}$_{0.25}$\textrm{MnO}$_{3}$
manganite.\cite{Booth}

Subramanian ed al.\cite{Subramanian} conclude from the properties of
\textrm{Tl}$_{2}$\textrm{Mn}$_{2}$\textrm{O}$_{7}$ that some of the
\textrm{Mn}$^{4+} $ electrons go into the \textrm{Tl} band, so that the
compound corresponds to \textrm{Tl}$_{2-x}^{3+}-$\textrm{Tl}$_{x}^{2+}%
$\textrm{Mn}$_{2-x}^{4+}$\textrm{Mn}$_{x}^{5+}$\textrm{O}$_{7}$, in agreement
with the model of Ventura and Alascio.\cite{Ventura1}

\subsection{The static resistivity and the magnetoresistance}

\begin{figure}[bh]
\includegraphics*[scale=0.80]{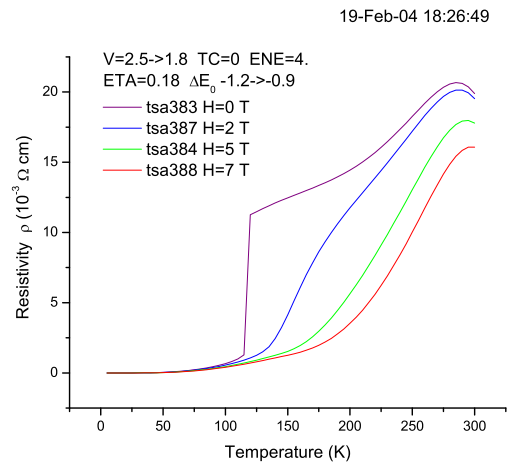}\caption{The resistivity as a
function of T for several magnetic fields, and for the system parameters
indicated in the figure.}%
\label{Fg02}%
\end{figure}

\begin{figure}[tbh]
\includegraphics*[scale=0.80]{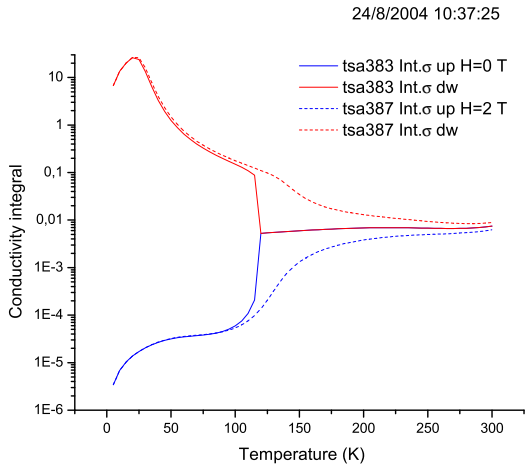}\caption{Values
proportional to $\sigma_{\sigma}\left(  T\right)  $ for up and down electrons
at $H=0$ T and $H=2$ T.}%
\label{Fg03a}%
\end{figure}

We employ Eq. (\ref{E3.4}) to calculate the resistivity for each spin
component, and we sum the two contributions to obtain the total conductivity.
In figure \ref{Fg02} we plot the resulting resistivity as a function of $T$
for the following magnetic fields: $B=0$ T, $2$ T, $5$ T, and $7$ T, and the
remaining system parameters are given in the figure. The values we calculated
are of the same order of those reported by Shimakawa ed al. \cite{Shimakawa0},
and there is a sharp increase in the resistance at the critical temperature
$T_{C}$; the increase becomes more gradual when the magnetic field increases.
In figure \ref{Fg03a} we plot quantities proportional to $\sigma_{\sigma
}\left(  T\right)  $ for the up and down electrons at both $H=0$ T and $H=2$ T.

\begin{figure}[tbh]
\includegraphics*[scale=0.80]{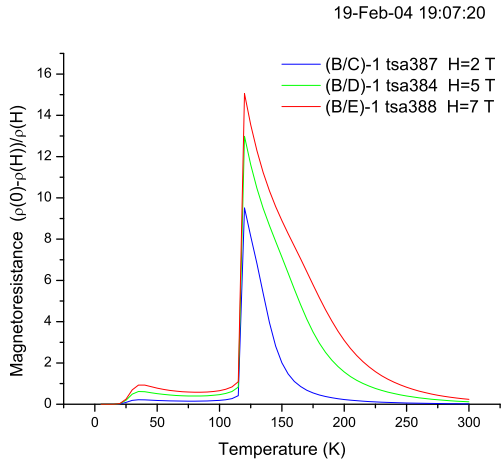}\caption{The
magnetoresistance as a function of T for several magnetic fields, and for
system parameters indicated in the figure.}%
\label{Fg03}%
\end{figure}

From the resistivity ad different magnetic fields we can calculate the
negative magnetoresistance ($\rho(B=0)-\rho(B))/\rho(B)$, plotted in figure
\ref{Fg03}. The value of the maximum near $T_{C}$ is close to that observed by
Cheong ed al.\cite{Cheong}, but the rise before the maximum is much steeper in
our calculation.

\subsection{The optical conductivity}

\begin{figure}[tbh]
\includegraphics*[scale=0.80]{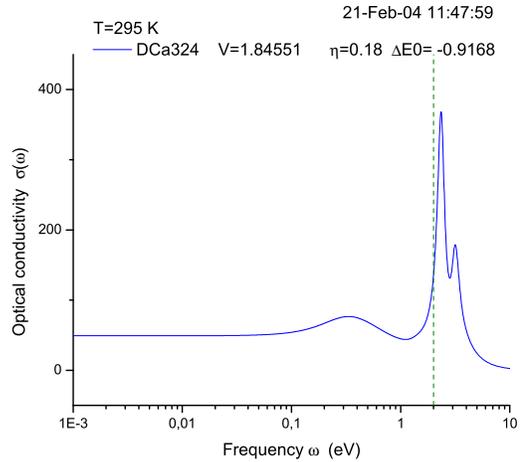}\caption{The optical
conductivity for system parameters indicated in the figure.}%
\label{Fg04}%
\end{figure}

\begin{figure}[tbh]
\includegraphics*[scale=0.80]{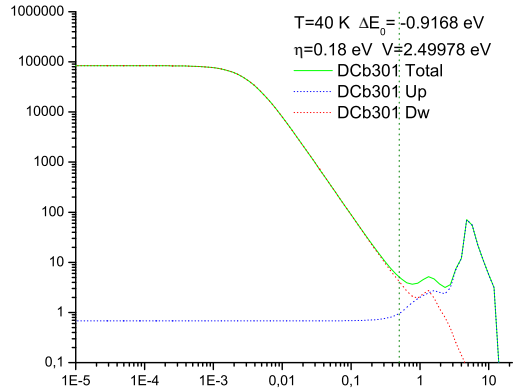}\caption{Contribution of the
two spin components to the optical conductivity at $T=40$ K for system
parameters indicated in the figure. The vertical line is located at 0.5 eV.}%
\label{Fg05}%
\end{figure}

We have employed Eq. (\ref{E3.1}) to calculate the optical conductivity
$\sigma\left(  \omega,T\right)  $ as the sum of the contribution
$\sigma_{\sigma}\left(  \omega,T\right)  $ of the two spin components. In the
measurements of the optical conductivity of Okamura ed al.\cite{Okamura} one
observes a strong peak close to $\omega=2$ eV at $T=295$ K. This type of
measurement is expected to depend on the value of the direct gap, which is
affected by the hybridization constant $V$.\ The static resistivity, on the
other hand, depends on the indirect gap, and the scattering mechanism at low
temperatures depends also on the hybridization, as discussed before. We have
then employed a temperature dependent value of $V$ to adjust these two
quantities, and we used $V\simeq1.85$eV at $T=295$ K . In figure \ref{Fg04} we
plot the the optical conductivity $\sigma\left(  \omega,T\right)  $ at
$T=295$~K, and we obtain a peak at the correct frequency. There are several
smaller peaks at lower frequencies, that have been asigned to optical
phonons.\cite{Okamura}

At low $T$ the two spin components make different contributions to
$\sigma\left(  \omega,T\right)  $, as shown in figure \ref{Fg05}. The
component with spin down corresponds to a metal, and the corresponding
$\sigma\left(  \omega,T\right)  $ describes the Drude peak of this metal, and
the figure shows that the metallic components are limited to below 0.5 eV, as
measured by Okamura ed al.\cite{Okamura}. Although the spectral density in
figure \ref{Fg01} does not give the direct gap, it is consistent with figure
\ref{Fg05}. The spin up component in figure \ref{Fg01} corresponds to a
semiconductor, and in figure \ref{Fg05} it shows a very small $\sigma\left(
\omega,T\right)  $ at low frequencies, that starts to increase at $\omega
\sim3$ eV. The different contribution of the two spin components can also be
understood by considering the sum rules\cite{RozenbergKK,BaeriswylGR,Foglio2}
of the two spin components of $\sigma_{\sigma}\left(  \omega,T\right)  $ as
two separate contributions, but we have not done a numerical analysis of this
interpretation. The two components give identical contributions to
$\sigma_{\sigma}\left(  \omega,T\right)  $ above $T_{C}$, because the two
bands are identical when the magnetization becomes zero.\begin{figure}[tbh]
\includegraphics*[scale=0.80]{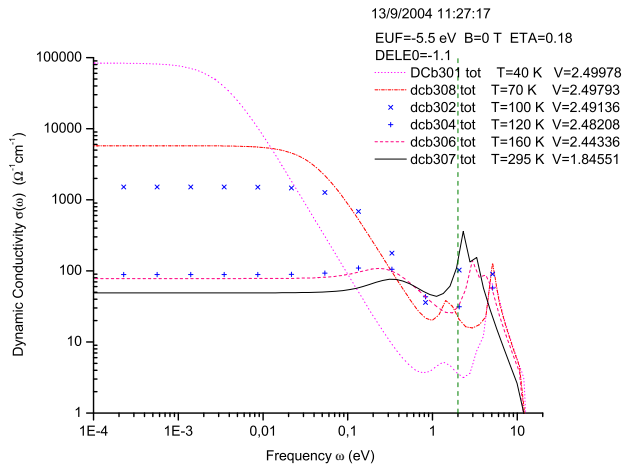}\caption{The optical
conductivity at low frequencies for several temperatures.The values of
\ $\sigma\left(  \omega,T\right)  $ when $\omega\rightarrow0$ give the
temperature dependence of the Drude peak}%
\label{Fg6a}%
\end{figure}

In figure \ref{Fg6a} we plot the optical conductivity for several
temperatures. At low frequencies one can see how the maximum of the Drude peak
at $\omega=0$ decreases and its width increases when $T$ increases.

\subsection{The thermopower and the magneto thermopower}

\begin{figure}[tbh]
\includegraphics*[scale=0.80]{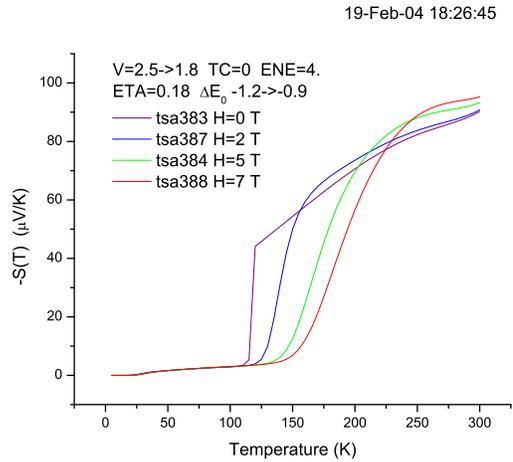}\caption{The thermopower for
system parameters indicated in the figure.}%
\label{Fg06}%
\end{figure}

\begin{figure}[tbh]
\includegraphics*[scale=0.80]{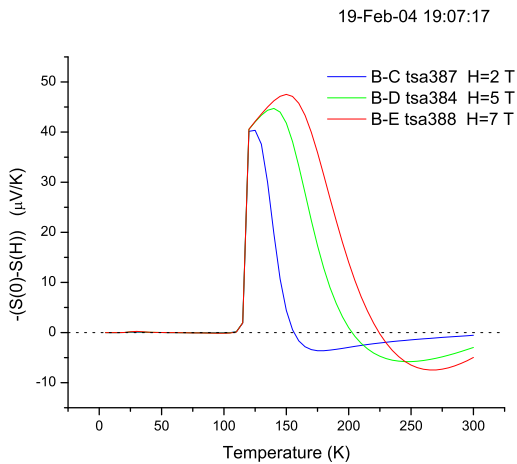}\caption{The magneto
thermopower for system parameters indicated in the figure.}%
\label{Fg07}%
\end{figure}

We have employed Eq. (\ref{E3.5}) to calculate the thermopower $S(H,T)$ of
\textrm{Tl}$_{2}$\textrm{Mn}$_{2}$\textrm{O}$_{7}$ within the model of Ventura
and Alascio, and in figure \ref{Fg06} we show the temperature dependence for
several magnetic fields. The plot agrees qualitatively with the experimental
results of Imai ed al.\cite{Imai} and at $H=0$ T it is approximately linear in
$T$ just below and above $T_{C}$ , but with different slopes. We employed Eq.
(\ref{E3.5}) to calculate $S(H,T)$, because the model is composed of two
independent subsystems. It is then straightforward to calculate the
magneto-thermopower, defined by $\Delta S(H)=S(H)-S(0)$, and in figure
\ref{Fg07} we plot our results, that show a semi quantitative agreement with
those in reference \onlinecite{Imai}: the magnitude of $\Delta S(H)$ is of the
same order, but the increase at $T_{C}$ is more abrupt than the one measured experimentally.

\section{\textbf{CONCLUSIONS}}

Asuming that the system is stoichiometric we have calculated the resistivity,
optical conductivity and thermopower as a function of temperature and magnetic
field of the model of \textrm{Tl}$_{2}$\textrm{Mn}$_{2}$\textrm{O}$_{7}$
introduced by Ventura and Alascio.\cite{Ventura1} Differently from other
studies, we have calculated these transport properties employing Kubo's
formula, that is directly related to the electronic GFs. To derive these GFs
we introduced Hubbard operators to describe the model, and used a treatment
previously employed to study \textrm{FeSi.\cite{Foglio1,Foglio2} }With the
dependence of resistivity and thermopower with magnetic field we have also
calculated the magneto resistance and the magneto thermopower.

We obtain a semiquantitative agreement with the experimental results by an
adequate choice of the system parameters, and we can conclude that the model
gives a fair description of all the calculated properties.

\begin{acknowledgments}
The authors would like to acknowledge financial support from the following
agencies: FAPESP and CNPq. They are grateful to B. R. Alascio and C. Ventura
for suggesting this study and for several interesting discussions.
\end{acknowledgments}

\appendix

\section{Atomic eigenstates}

In table \ref{T1} we give the atomic eigenstates $\mid j,\nu,r\rangle$ of
$H_{j}$ (cf. Eq. (\ref{E2.9})) as a function of the eigenstates for $V=0$. To
abbreviate we use $E_\pm=E_{\pm1/2}$, $\varepsilon_\pm=E_{\pm1}-\mu$,
$\varepsilon_\pm^{0}=E_{0,\pm}^{a}-\mu$, and $\varepsilon_{2}^{0}%
=\varepsilon_{+}^{0}+\varepsilon_{-}^{0}$ as well as the following energy
expressions that appear often in the formulas%
\begin{align*}
\varepsilon_{m\pm}  &  =\left(  \varepsilon_\pm-E_\pm-\varepsilon_{\pm
}^{0}\right)  /2\\
\varepsilon_{s\pm}  &  =\left(  \varepsilon_\pm+E_\pm+\varepsilon_{\pm
}^{0}\right)  /2\\
r_\pm &  =\sqrt{\left(  \varepsilon_{m\pm}\right)  ^{2}+\left\vert
V\right\vert ^{2}}%
\end{align*}
The coefficients of the eigenfunctions in table \ref{T1} are obtained from%
\[
tg\phi_\pm=\pm tg\ \psi_\pm=\frac{V^{\ast}}{\varepsilon_{m\pm}+r_\pm},
\]
and we conventionally use $\cos\phi_\pm>0$ and $\cos\psi_\pm>0$\ to
specify the sign of the eigenfunctions.\

\begin{table}[ptb]
\par%
\begin{tabular}
[c]{|l|l|l|l|}\hline
$\left\vert n,r\right\rangle $ & Eigenstate & $S_{z}$ & $\varepsilon_{r}%
=E_{r}-n_{r}\mu$\\\hline\hline
$\left\vert 0,1\right\rangle $ & $\left\vert -\frac{1}{2},0\right\rangle $ &
$-1/2$ & $\varepsilon_{1}=E_{-}$\\\hline
$\left\vert 0,2\right\rangle $ & $\left\vert +\frac{1}{2},0\right\rangle $ &
$+1/2$ & $\varepsilon_{2}=E_{+}$\\\hline
$\left\vert 1,3\right\rangle $ & $C\phi_{-}\left\vert -1,0\right\rangle
-S\phi_{-}\left\vert -\frac{1}{2},\downarrow\right\rangle $ & $-1$ &
$\varepsilon_{3}=\varepsilon_{s-}+r_{-}$\\\hline
$\left\vert 1,4\right\rangle $ & $C\phi_{+}\left\vert +1,0\right\rangle
-S\phi_{+}\left\vert +\frac{1}{2},\uparrow\right\rangle $ & $+1$ &
$\varepsilon_{4}=\varepsilon_{s+}+r_{+}$\\\hline
$\left\vert 1,5\right\rangle $ & $S\phi_{-}\left\vert -1,0\right\rangle
+C\phi_{-}\left\vert -\frac{1}{2},\downarrow\right\rangle $ & $-1$ &
$\varepsilon_{5}=\varepsilon_{s-}-r_{-}$\\\hline
$\left\vert 1,6\right\rangle $ & $S\phi_{+}\left\vert +1,0\right\rangle
+C\phi_{+}\left\vert +\frac{1}{2},\uparrow\right\rangle $ & $+1$ &
$\varepsilon_{6}=\varepsilon_{s+}-r_{+}$\\\hline
$\left\vert 1,7\right\rangle $ & $\left\vert -\frac{1}{2},\uparrow
\right\rangle $ & $0$ & $\varepsilon_{7}=E_{-}+\varepsilon_{+}^{0}$\\\hline
$\left\vert 1,8\right\rangle $ & $\left\vert +\frac{1}{2},\downarrow
\right\rangle $ & $0$ & $\varepsilon_{8}=E_{+}+\varepsilon_{-}^{0}$\\\hline
$\left\vert 2,9\right\rangle $ & $\left\vert -1,\downarrow\right\rangle $ &
$-3/2$ & $\varepsilon_{9}=\varepsilon_{-}+\varepsilon_{-}^{0}$\\\hline
$\left\vert 2,10\right\rangle $ & $\left\vert +1,\uparrow\right\rangle $ &
$+3/2$ & $\varepsilon_{10}=\varepsilon_{+}+\varepsilon_{+}^{0}$\\\hline
$\left\vert 2,11\right\rangle $ & $C\psi_{-}\left\vert -1,\uparrow
\right\rangle -S\psi_{-}\left\vert -\frac{1}{2},\uparrow\downarrow
\right\rangle $ & $-1/2$ & $\varepsilon_{11}=\varepsilon_{3}+\varepsilon
_{+}^{0}$\\\hline
$\left\vert 2,12\right\rangle $ & $C\psi_{+}\left\vert +1,\downarrow
\right\rangle -S\psi_{+}\left\vert +\frac{1}{2},\uparrow\downarrow
\right\rangle $ & $+1/2$ & $\varepsilon_{12}=\varepsilon_{4}+\varepsilon
_{-}^{0}$\\\hline
$\left\vert 2,13\right\rangle $ & $S\psi_{-}\left\vert -1,\uparrow
\right\rangle +C\psi_{-}\left\vert -\frac{1}{2},\uparrow\downarrow
\right\rangle $ & $-1/2$ & $\varepsilon_{13}=\varepsilon_{5}+\varepsilon
_{+}^{0}$\\\hline
$\left\vert 2,14\right\rangle $ & $S\psi_{+}\left\vert +1,\downarrow
\right\rangle +C\psi_{+}\left\vert +\frac{1}{2},\uparrow\downarrow
\right\rangle $ & $+1/2$ & $\varepsilon_{14}=\varepsilon_{6}+\varepsilon
_{-}^{0}$\\\hline
$\left\vert 3,15\right\rangle $ & $\left\vert -1,\uparrow\downarrow
\right\rangle $ & $-1$ & $\varepsilon_{15}=\varepsilon_{-}+\varepsilon_{2}%
^{0}$\\\hline
$\left\vert 3,16\right\rangle $ & $\left\vert +1,\uparrow\downarrow
\right\rangle $ & $+1$ & $\varepsilon_{16}=\varepsilon_{+}+\varepsilon_{2}%
^{0}$\\\hline
\end{tabular}
\par
\caption[TABLE I]{ \ The sixteen eigenstates $\mid n,r\rangle$ of
$\mathcal{H}$ are given as a function of the eigenstates in the absence of
hybridization, together with their eigenvalues $\varepsilon_{n,r}=E_{n,r}%
-n\mu$, where $E_{n,r}$ is the energy of the state $\mid n,r\rangle$. To
abbreviate we used $C\phi_\pm= $cos$\phi_\pm$, $S\phi_\pm=$sen$\phi
_\pm$, {\ }$C\psi_\pm= $cos$\psi_\pm$ and $S\psi_\pm=$sen$\psi_\pm$.
}%
\label{T1}%
\end{table}



\end{document}